\documentclass[conf]{new-aiaa}
\usepackage[utf8]{inputenc}
\usepackage{graphicx}
\usepackage{amsmath}
\usepackage[version=4]{mhchem}
\usepackage{siunitx}
\usepackage{longtable,tabularx}
\usepackage{xcolor}
\usepackage{bm}

\usepackage{fancyhdr}
\usepackage{blindtext}
\usepackage{eso-pic, rotating, graphicx}
\AddToShipoutPicture{\put(30,300){\rotatebox{90}{\scalebox{1.25}{71st JANNAF Propulsion Meeting, Oklahoma City, OK, 6-10 May 2024}}}}

\title{Deep operator learning-based surrogate models for aerothermodynamic analysis of AEDC hypersonic waverider}
\author[1]{Khemraj Shukla}
\author[2]{Jasmine Ratchford}
\author[3]{Luis Bravo}
\author[1]{Vivek Oommen}
\author[3]{Nicholas Plewacki}
\author[3]{Anindya Ghoshal}
\author[1]{George Karniadakis} 

\affil[1]{Applied Mathematics Department, Brown University, Providence, Rhode Island, USA}
\affil[2]{Carnegie Mellon University Software Engineering Institute, Pittsburgh, PA  US}
\affil[3]{DEVCOM Army Research Laboratory, Aberdeen Proving Ground, MD, USA}

\begin{document}
\maketitle

\begin{abstract}
Neural networks are universal approximators that traditionally have been used to learn a map between function inputs and outputs. However, recent research has demonstrated that deep neural networks can be used to approximate operators, learning function-to-function mappings.
Creating surrogate models to supplement computationally expensive hypersonic aerothermodynamic models in characterizing the response of flow fields at different angles of attack (AoA) is an ideal application of neural operators. We investigate the use of neural operators to infer flow fields (volume and surface quantities) around a geometry based on a 3D waverider model based on experimental data measured at the Arnold Engineering Development Center (AEDC) Hypervelocity Wind Tunnel Number 9. We use a DeepONet neural operator which consists of two neural networks, commonly called a branch and a trunk network. The final output is the inner product of the output of the branch network and the output of the trunk net.
Because the flow field contains shocks across the entire volume, we conduct a two-step training approach of the DeepONet that facilitates accurate approximation of solutions even in the presence of discontinuities. We train various DeepONet models to understand and predict pressure $(p)$, density $(\rho)$, velocity $(\bm{u})$, heat flux $(Q_w)$, and total shear stress $(\tau_{w})$ for the AEDC waverider geometry at Ma=7.36 across AoA that range from $-10^{\circ}$ to $10^{\circ}$ for surface quantities and from $-14^{\circ}$ to $14^{\circ}$ for volume quantities. 

\end{abstract}

\section{Introduction}
Neural networks are well known as universal function approximators that can solve regression problems, mapping input data to output data. Recently, a change in perspective, initiated by the seminal paper on the deep operator network or DeepONet (Lu et al., 2021; Lu et al., 2019), demonstrated that neural networks can also act as operators, mapping between two functional spaces. In contrast to other physics-informed neural networks (PINNs) as described in Raissi et al. (2019) that learn fixed mappings for specific conditions, neural operators learn parametric function mappings. This feature allows for real-time applications such as forecasting, design, autonomy, and control.

DeepONets have the capability to handle multi-fidelity or multi-modal input \cite{de2022bi, howard2022multifidelity, lu2022multifidelity, jin2022mionet, zhu2022reliable} within one network, while using an independent network to represent the output space, such as in space-time coordinates or continuous parametric space. In a sense, DeepONets can serve as surrogates akin to reduced order models (ROMs)  \cite{hesthaven2018non, hesthaven2016certified, benner2017model, williams2015data, chiavazzo2014reduced, lieberman2010parameter, bui2008model, benner2015survey, amsallem2015design, carlberg2008compact, choi2020gradient}. However, unlike ROMs, they exhibit over-parameterization, leading to enhanced generalizability and noise robustness, a distinction elaborated in the recent work by Kontolati et al. (2022). Neural operators are a valuable modeling tool in engineering: the capacity to substitute highly intricate and computationally intensive multiphysics systems with neural operators capable of delivering functional outputs in real-time. Figure \ref{fig: DeepONet_diagram} is an architectural schematic of a DeepONet. The figure includes labels to illustrate the commonly adopted nomenclature used to describe DeepONets components.

In the present work, we investigate the possibility of using DeepONets for prediction of the flow fields over different angles of attack (AoAs), an idea that has not been explored before. In particular, we focus on constructing computational fluid dynamic (CFD) surrogates for the 3D Arnold Engineering Development Center (AEDC) Hypervelocity Wind Tunnel Number 9 \cite{hypersonic_waverider} Waverider (hereafter called the AEDC waverider) at different AoAs under hypersonic flow conditions. Conventionally, such optimization processes rely on computationally intensive compressible flow numerical solvers to accurately model flow fields around intricate geometries. Replacing full CFD simulations with acceptable accuracy surrogate models can significantly accelerate the optimization loop by removing the time-consuming aspects inherent to numerical solvers.

With the significant advancement in computational power, Deep Neural Network (DNN) tools have gained much attention for serving as accurate surrogate models in a broad spectrum of scientific disciplines \cite{zhang2021multi,zhiwei2020non,renganathan2021enhanced}, and in other applications such as time-series classifications \cite{xing2022selfmatch, xiao2021rtfn} and as engineering aids, such as in diagnosing bearing faults \cite{mishra2022intelligent, mishra2022self, mishra2022fault}. The DNN approach can be readily trained for numerous input design variables to predict the cost function of the optimization loop. Du \emph{et al.} \cite{du2021rapid} trained a feed-forward DNN to receive airfoil shapes and predict drag and lift coefficients. They also used RNN models for estimating the pressure coefficient. The optimal airfoil design determined using the surrogate model was compared with an airfoil design obtained with a CFD-based optimization process \cite{du2021rapid}. Hao \emph{et al.} \cite{pmlr-v202-hao23c} provides a comparative study of neural operator learning methods for flow field prediction around airfoils. Liao \emph{et al.} \cite{liao2021multi} designed a surrogate model using a multi-fidelity Convolutional Neural Network (CNN) with transfer learning. This learning method transfers the information learned in a specific domain to a similar field. The low-fidelity samples are taken as the source, and the high-fidelity ones are assigned as targets. Tao and Sun \cite{tao2019application} introduced a Deep Belief Network (DBN) to be trained with low-fidelity data. The trained DBN was later combined with high-fidelity data using regression to create a surrogate model for shape optimization. Existing surrogate models for shape optimization are all trained to predict lift, drag, or pressure coefficients. For example, Zhao \emph{et al.}\cite{zhao2023learning} uses a DeepONet to learn the mapping from iced airfoil geometries to its aerodynamic coefficients. In contrast, the flow field around the aerodynamic shape is not inferred. Prior works have also investigated the capabilities and limitations of the different neural operators in a variety of benchmark cases in \cite{lu2022comprehensive}. The recent Geo-FNO \cite{li2022fourier} and CORAL \cite{serrano2023operator} proposes neural operator-based models that are capable of learning solutions of PDEs on general geometries. However, both of these studies overlooked viscous forces in lift and drag calculations, reducing the realism of their results. Here, we construct a surrogate model that predicts the viscous flow field around the AEDC waverider at hypersonic conditions using DeepONet. The objective is to develop a CFD surrogate at hypersonic flow condition that can infer the flow field at unseen AoAs. The surrogate model is constructed using DeepONet and is trained using high-fidelity CFD simulations of a hypersonic flow regime. 
This analysis is preceded by a discussion of early related work on the High-speed Army Reference Vehicle (HARV) \cite{harv_2022} at supersonic and hypersonic speeds that provided early validation of our approach. The novel work conducted and summarized in this work include:
\begin{itemize}
     \item We created a surrogate model for a 3-dimensional, non-trivial flow and geometry based on DeepONets, offering an efficient and cost-effective alternative to the expensive CFD solver.
     \item We compared the basic DeepONet approach as described in \cite{lu2019deeponet}, along with the two-step DeepONet methodology outlined in \cite{lee2023training}, to predict both surface and volume flow fields quantities.
     \item We showed that the two-step approach accurately capture shock behavior and suggest that it enhances the interpretability of the surrogate model.
\end{itemize}


\section{Computational Methodology}

\subsection{Methodology}
%
\subsubsection{Brief Review of DeepONets}
\begin{figure}
\centering
    \includegraphics[scale=0.4]{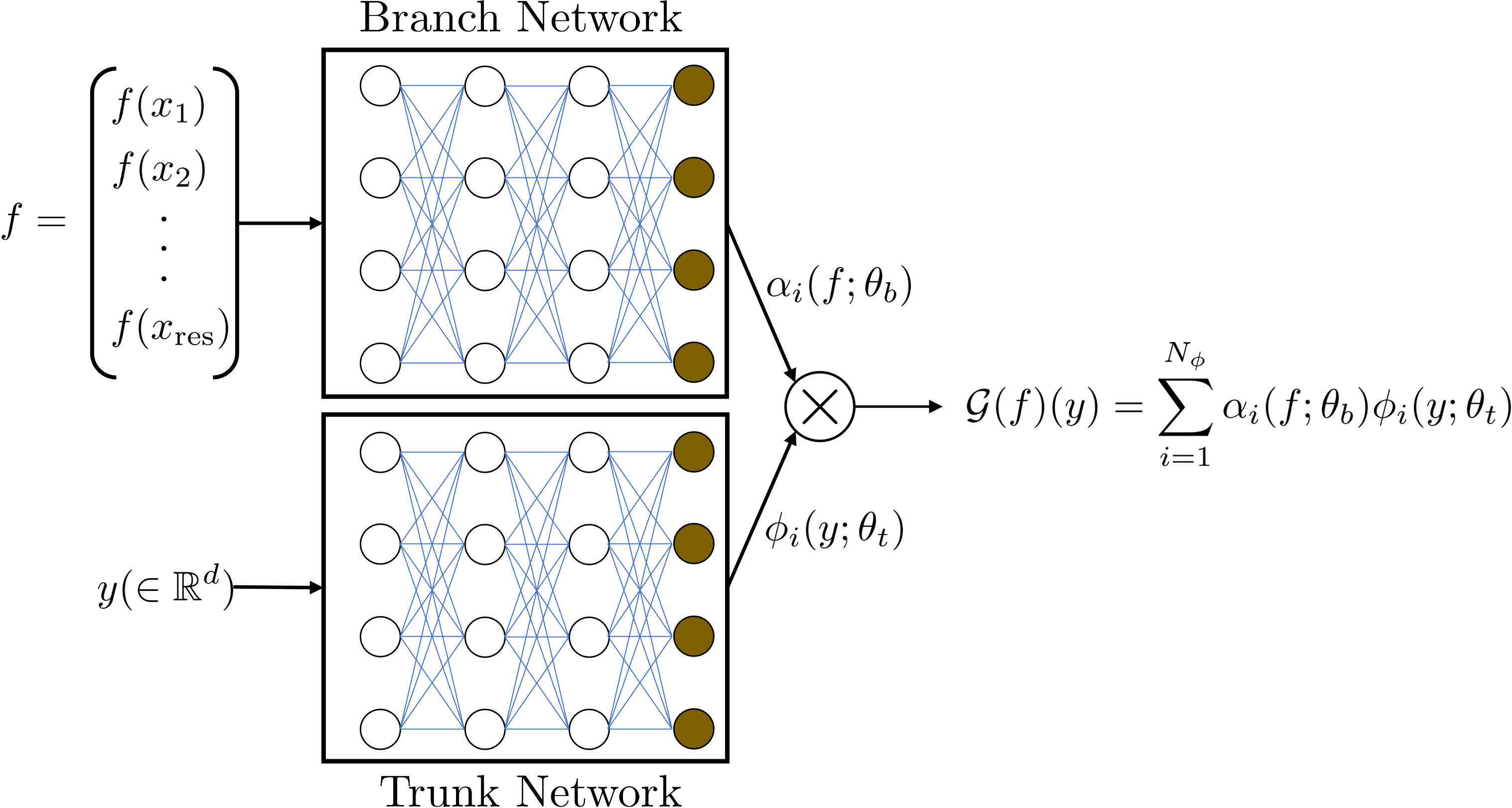}
  \caption{\textmd{A schematic representation of a DeepONet that is trained to learn the mapping from the input function $f$ to the output function $\mathcal{G}(f)(y)$, evaluated at $y$. DeepOnet consists of a branch and a trunk network.}}
  \label{fig: DeepONet_diagram}
\end{figure}

Neural operators are neural network models developed based on the universal operator approximation theorem \cite{chen1995universal}.  The neural operators learn the mapping between spaces of function and directly learn the underlying operator from the available training data. DeepONets \cite{lu2021learning} and Fourier Neural Operators (FNO) \cite{li2020neural} are the two popular neural operators extensively used for solving a wide spectrum of problems in diverse scientific areas. A schematic representation of DeepONet architecture is given in \autoref{fig: DeepONet_diagram}. A DeepONet consists of a branch network that encodes the input function and a trunk network that learns a collection of basis functions. The DeepONet output is computed by taking the inner product between the branch and trunk network outputs.  
\subsubsection{Training "vanilla" DeepONets} \label{two_step}
 We began by training separate DeepONet models to learn separate flow fields (e.g., density ($\rho$)) 
 given an angle of attack ($\xi_g$) for  specified geometry parameters using training data generated from CFD simulations. The trunk network learns a collection of basis (${\phi}$) as functions of spatial coordinates, and the branch network learns the corresponding coefficients (${\alpha}$) as a function of the angle of attacks. The DeepONet output is defined as
\begin{equation}\label{dnet_eq}
    \mathcal{G}^{q}(\xi_g)(x,y) = \sum_{i=1}^{N_{\phi}} \alpha_i(\xi_g; \theta_b^q) \phi_i(x,y; \theta_t^q) \quad  q \in \{ \rho, \tau_w, Q_w\}.
\end{equation}



When originally constructed, DeepONet training consisted of a single training step during which both the branch and trunk net were trained simultaneously. We call this "vanilla" DeepONet. Due to the high Mach number of the scenarios that we investigate, shocks are present in the simulation output used for training. These shocks contribute to slow convergence of DeepONet training when both the branch and trunk networks are trained in the "vanilla" DeepONet fashion. 
 
\subsubsection{Two-step Training of DeepONet} 
In order to address the slow convergence of the DeepONet due to the presence of shocks,  we implement a a two-step training approach \cite{lee2023training} to train the DeepONet. Assuming that the number of output "sensors" ($m_y$) is larger than the width of the final layer ($N_\phi$), we can conduct the two-step training method for solving \autoref{dnet_eq}.

Step 1. In the first step, the trunk network $\phi$ is evaluated for the following minimization problem:
\begin{align}\label{step1}
\min _{\theta_t, A} \mathcal{L}(\theta_t, A):=\|{\phi}(\theta_t) A-\boldsymbol{U}\|_{p, p}^p \quad \text { where } \quad A \in \mathbb{R}^{(N+1) \times K} ,
\end{align}
where $\theta_t$ represents the trainable parameters of the trunk network, $A$ is a trainable matrix that represents the branch network, $N$ is number of training samples in branch network and $U$ is labeled data. If the optimal solution is $\left(\theta_t^*, A^*\right)$, and ${\phi}\left(\theta_t^*\right)$ is full rank,  we can set $T^*=$ $\left(R^*\right)^{-1}$, where $R^*$ is obtained from a QR-factorization of $\phi\left(\mu^*\right)$, i.e., $Q^* R^*=$ $\phi\left(\theta_t^*\right)$. The trunk network is then fully determined as $\hat{\phi}\left(\cdot ; \theta_t^*, T^*\right) = T^T\phi\left(\cdot ; \theta_t^*\right) $.

Step 2. The second step consists of training the branch network $\alpha(\theta_b)$ to fit $R^* A^*$. Specifically, we consider the optimization problem of
\begin{align}\label{step_2}
\min _{\theta_b}\left\|\alpha(\theta_b)-R^* A^*\right\|_{2,2}^2 .
\end{align}
Assuming $\theta_b^*$ to be an optimal solution for the branch network, the fully trained branch network is given by $\boldsymbol{c}\left(\cdot ; \theta_b^*\right)$. 

The first step replaces the use of the branch network from \autoref{dnet_eq} to the corresponding value matrix $A$. Because the trunk loss function is convex with respect to $A$ (assuming $p \geq 1$), this method avoids convergence challenges due to branch network nonlinearity and nonconvexity. In addition, this training mechanism modifies the number of trainable parameters from $|\theta_t|+|\theta_b|$ to $|\theta_t|+|A|$. This is a dimensional reduction of the optimization problem when $|A|=(N+1) K<|\theta|$.


\section{Results and Discussions}

\subsection{High-speed Army Reference Vehicle (HARV) - Supersonic - Hypersonic study}\label{volume_results_harv}
\begin{figure}
\includegraphics[width=\textwidth]{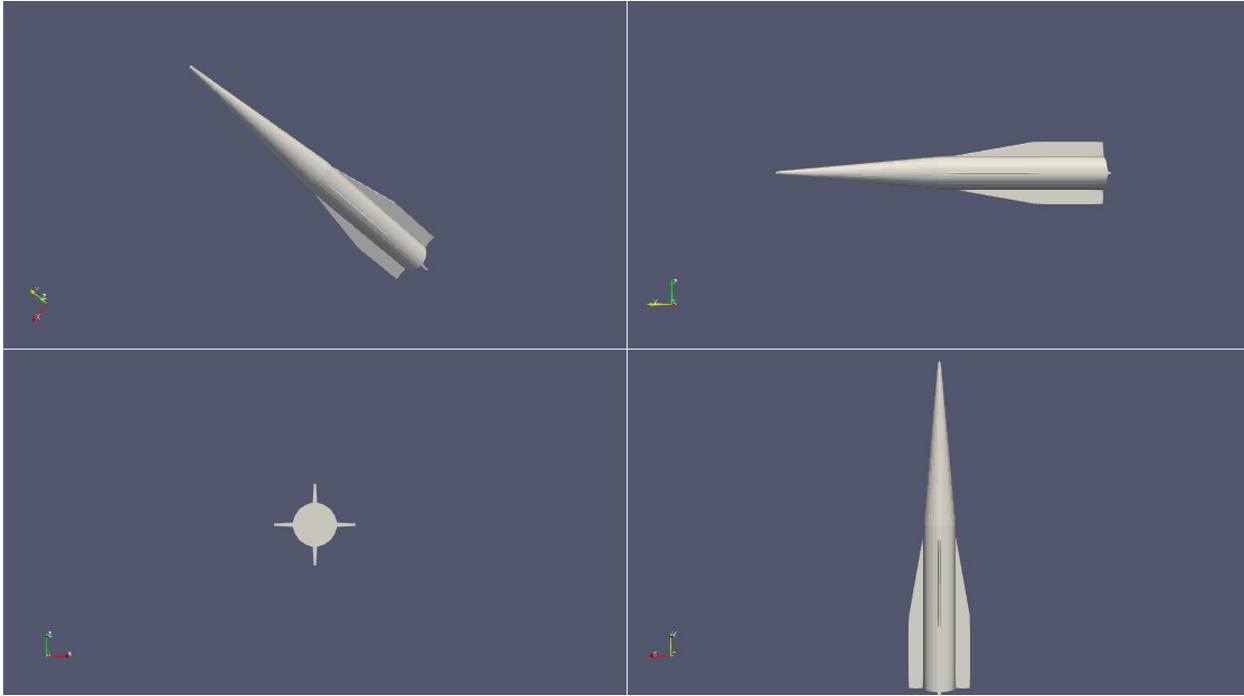}
\caption{Geometry of High-speed Army Reference Vehicle (HARV)}
  \label{fig:harv}
\end{figure}

To demonstrate DeepONet's effectiveness across various paremeterization and geometry settings, we constructed it for the HARV \cite{harv_2022} shown in \autoref{fig:harv}. In this computational experiment, the branch net takes Mach number as input, ranging from 5 to 20 with increments of 1 during training. Inference is then conducted for Mach numbers between 5.5 and 19.5, with increments of 1. At such high Mach numbers, the flow field exhibits shocks characterized by high gradients. To accurately capture these shocks, we assign greater weights to the flow field at the location of the shocks during DeepONet training. The shock locations were determined by calculating the flow field gradients using a Sobel filter in both the x and y directions. The convolution operation for density $(\rho)$ is expressed as follows:

\begin{align} \label{sobel}
\begin{aligned}
& \mathbf{G}_x=\left[\begin{array}{lll}
+1 & 0 & -1 \\
+2 & 0 & -2 \\
+1 & 0 & -1
\end{array}\right] * \mathbf{\rho} \quad \text { and } \quad \mathbf{G}_y=\left[\begin{array}{ccc}
+1 & +2 & +1 \\
0 & 0 & 0 \\
-1 & -2 & -1
\end{array}\right] * \mathbf{\rho}, \\
& \mathbf{G}=\sqrt{\mathbf{G}_x^2+\mathbf{G}_y{ }^2},
\end{aligned}
\end{align}
where $G_x$ and $G_y$ are gradient of $\rho$ in $x$ and $y$ direction, respectively. $G$ represents the total gradient for the $\rho$.

\begin{figure}
\includegraphics[width=\textwidth]{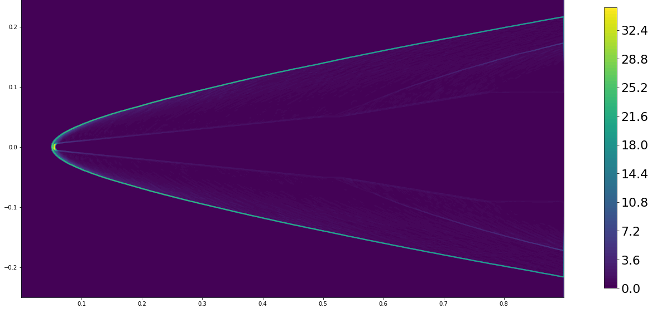}
\caption{Gradient of the $G = \nabla \rho$ computed using \autoref{sobel} for Ma=20. It is to be noted that the high gradient region clearly stands out and this will be used for informing the DeepONet during training phase.}
  \label{fig:harv_grad}
\end{figure}

\begin{figure}
\includegraphics[width=\textwidth]{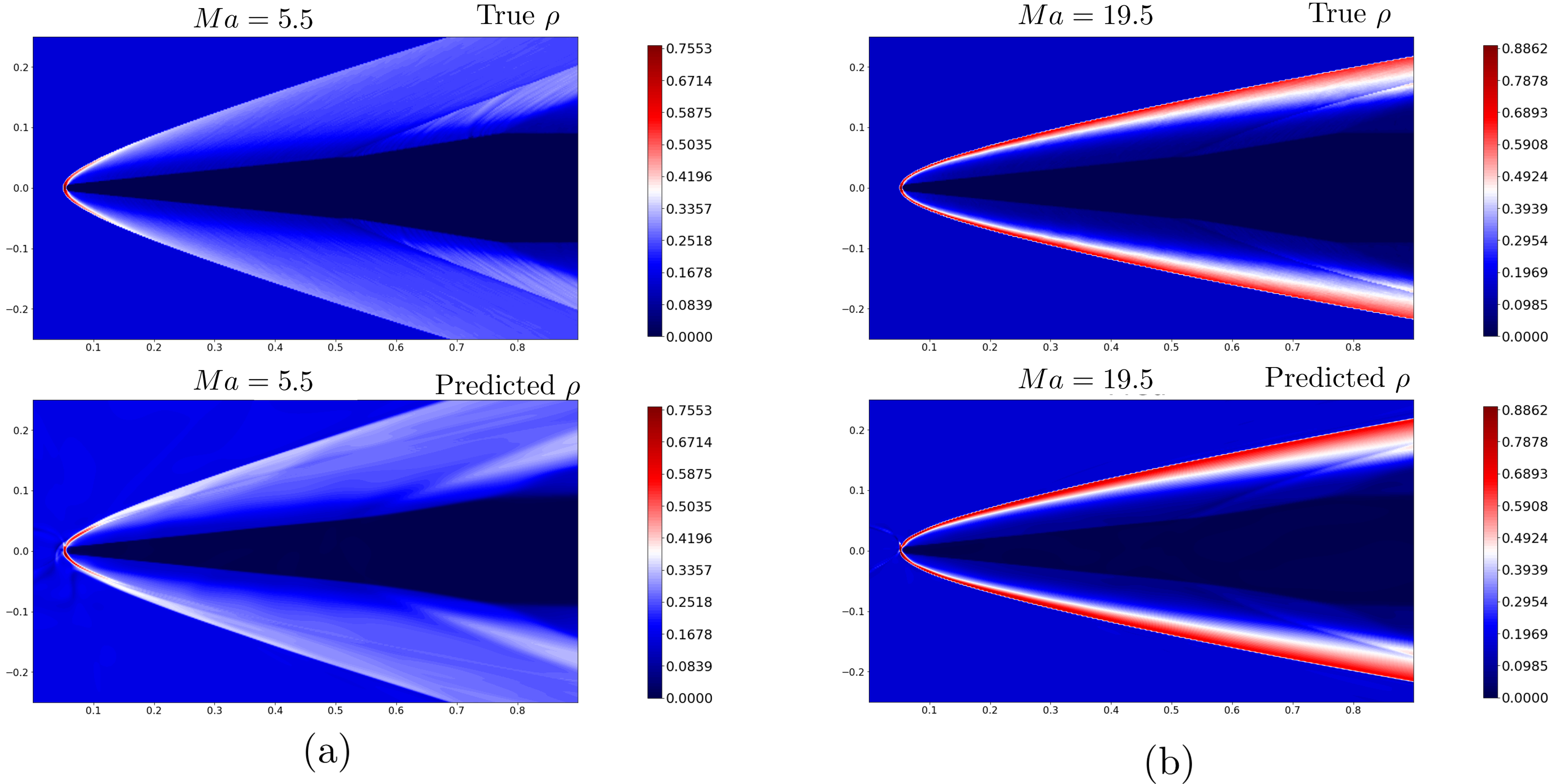}
\caption{True and predicted $\rho$ by DeepONet at (a) $Ma=5.5$ and (b) $Ma=19.5$. The global relative $L_2$-error between actual and predicted $\rho$ is 0.2\% and 0.6\%, respectively.}
  \label{fig:harv_rho}
\end{figure}

In \autoref{fig:harv_grad} we display the gradient $G$ calculated using the \autoref{sobel} method for Mach number 20 to demonstrate the validity and accuracy of the filtering process. The shock location is clearly visible, characterized by a region of high gradient and the Sobel filter captures it accurately. To incorporate this gradient information into the DeepONet, we formulate the loss function as follows:
\begin{align}\label{shock_loss}
\mathcal{L}=\sum_i \lambda\left(x_i, y_i\right)\left(\rho\left(x_i, y_i\right)-\hat{\rho}\left(x_i, y_i\right)\right)^2,
\end{align}
where $\rho$ and $\hat{\rho}$ is actual and predicted density. $\lambda$ is weighting coefficients and computed by using the $G=\nabla \rho$ and given as 
\begin{align}\label{lambda}
\lambda\left(x_i, y_i\right)=1+\epsilon \frac{\left|G\left(x_i, y_i\right)\right|}{\max (G)}
\end{align}
with $\epsilon=1$ but can be tuned.
In \autoref{fig:harv_rho} we present $\rho$ predicted by DeepONet at $Ma=5.5$ (supersonic regime) and $Ma=19.5$ (hypersonic regime). \autoref{fig:harv_rho} clearly shows a very good agreement between actual and predicted $\rho$. The global $L_2$-norm of relative error is 0.2\% and 0.6\% for $Ma=5.5$ and $Ma=19.5$, respectively.

\subsection{Hypersonic Waverider Study for Surface Quantities} \label{surface_results}

We generated hypersonic aerothermodynamic data necessary to train the DeepONet surrogate model of the  the AEDC waverider geometry using the US3D commercial CFD package. US3D is a state-of-the-art analysis tool developed as a collaborative effort between NASA Ames, the University of Minnesota, and VirtusAero, Inc. This code is massively parallel using the Message Passing Interface (MPI) libraries and is deployed on the Department of Defense (DoD) High-Performance Computing (HPC) system Warhawk, which we used in this study. US3D solves the compressible Navier-Stokes equations on an unstructured finite-volume mesh with high-order, low-dissipation fluxes. The solver has been tailored to excel at the complex evaluation of hypersonic flows including strong shocks, shock boundary layer interactions, and plasma dynamics, and has well-demonstrated accuracy for applied hypersonic configurations \cite{US3D}.

For this study we set the free stream and surface boundary conditions to be consistent with the reported experimental conditions as follows: $\rho_{\inf} = 0.5644 \text{kg/m$^3$}$, $T_{\inf} = 72.77~\text{K}$, and $v_{\inf} = 1279.25~\text{m/s}$ with Mach number of $7.36$. The surface temperature of the AEDC waverider is isothermal and held at 300K based on experimental conditions. We modeled turbulence using the classical Menter-SST Reynolds Averaged Navier Stokes (RANS) formulation (with a vorticity source term) along with $5$ species of chemical kinetics to handle the non-equilibrium chemistry. The dataset used to train the surrogate model is comprised of 21 simulations run with the AoAs varied between $-10^\circ$ and $+10^\circ$ by $1$-degree increments. This provides a wide range of aerothermodynamic loading as reported in the AEDC wind tunnel. We used the meshing software LINK3D to create the grid, which consisted of 50.4 million cells with wall spacing producing y+ values well below one. In addition, the wake region behind the waverider was excluded, and the fluid domain ends at the rear of the vehicle. We ran the simulations to $20+$ flow through times to ensure that shock structures and boundary layers are well established and that the flow solution is stable. Herein, we demonstrate the application of DeepONet for approximating the heat flux ($Q_w$) and shear stress ($\tau_y$) fields around an AEDC waverider. The dataset consists of $Q_w$ and $\tau_{y}$ fields at the surface of the waverider geometry at each of 21 AoAs in the dataset as mentioned above. Of the 21 AoAs simulated, 13 were used for training the DeepONet surrogate and 9 were held for testing. The input to the DeepONet trunk net consisted of approximately 250K surface grid node locations.

\begin{figure}
\centering
\includegraphics[trim={0cm 0cm 0 0},clip, width=\textwidth]{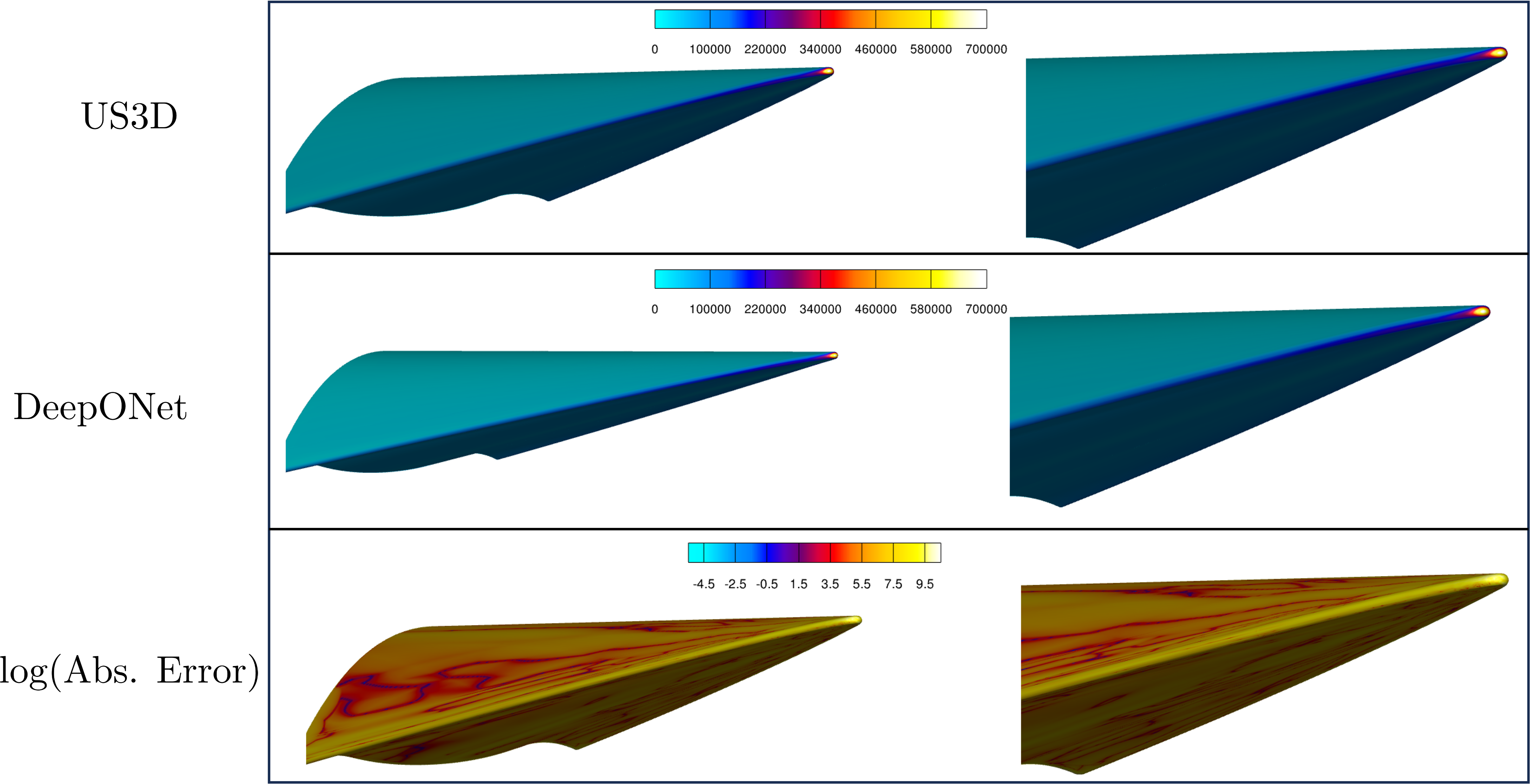}
  \caption{Heat flux ($Q_w$) distributions on the surface of the AEDC waverider. We compare the fields simulated by the US3D solver against those predicted by the surrogate DeepONet for an unseen 2$^\circ$ angle of attack. }
  \label{fig:3D_comparison}
\end{figure}

Next, we compare $Q_w$ field simulated by the US3D solver with the surrogate 3D DeepONet's predictions across the entire surface of the waverider. In \autoref{fig:3D_comparison}, we present the heat flux field at the surface of the AEDC waverider at $2^\circ$ AoA. The right column represents a zoomed-in view of the leading edge of the waverider to better visualize the quality of the surrogate 3D DeepONet prediction at the region where the variance of the fields is the largest. For this test sample we observe that the maximum absolute errors for heat flux and shear stress are 9.7\% and 4.1\% respectively. 

\subsection{Hypersonic Waverider Study for Volume Quantities}\label{volume_results_aedc}
 We employ the two-step method outlined in \autoref{two_step} to train a DeepONet for the flow fields across the entire volume. The data generation process mirrors that of the scenario detailed in \autoref{surface_results}. However, in this instance, we used a larger dataset encompassing angles of attack with 1-degree increments ranging from $-14^\circ$ to $+14^\circ$, resulting in a total of 29 AoAs. Of these 29 angles, 20 were used for training and 9 were set aside for testing. Subsequently, we proceed to compare the density ($\rho$) field simulated by the US3D solver with the predictions generated by the surrogate 3D DeepONet across the entire domain. Because the input to the volume DeepONet is the set of approximately 50.76 M grid node locations, it necessitates a deeper neural network to converge.

\begin{figure}
\includegraphics[scale=0.395]{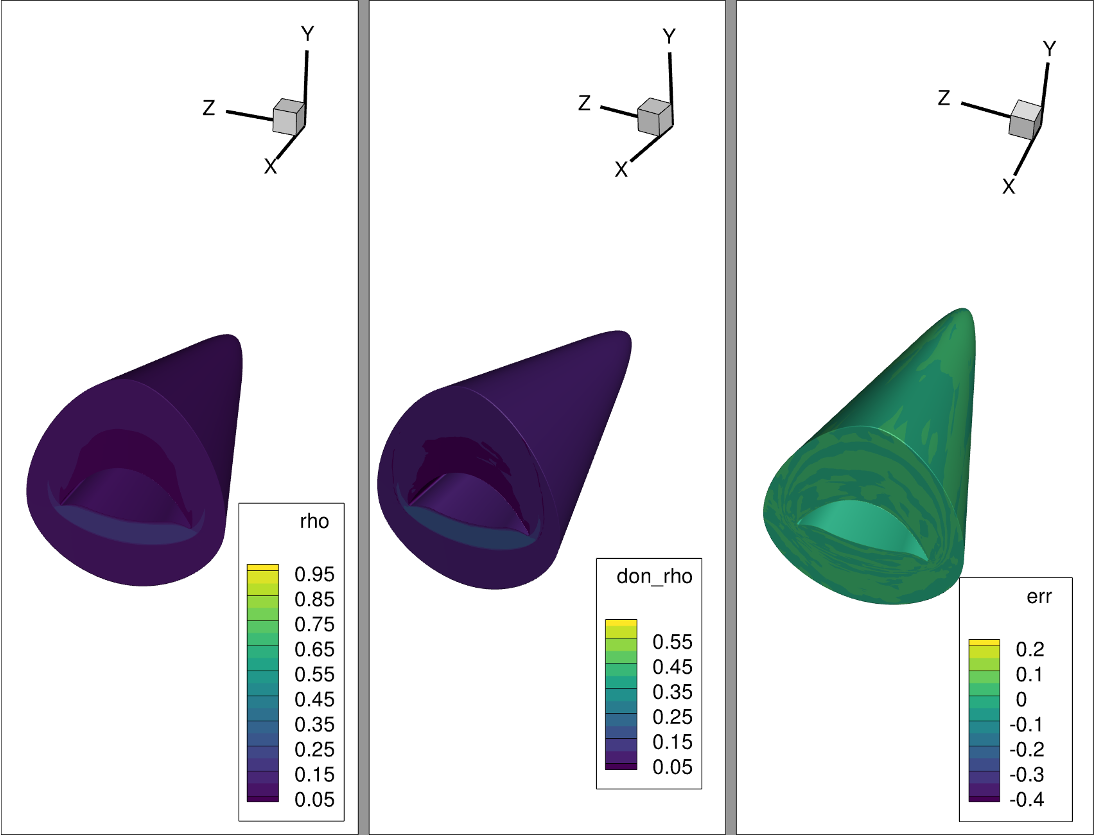}
\includegraphics[scale=0.38]{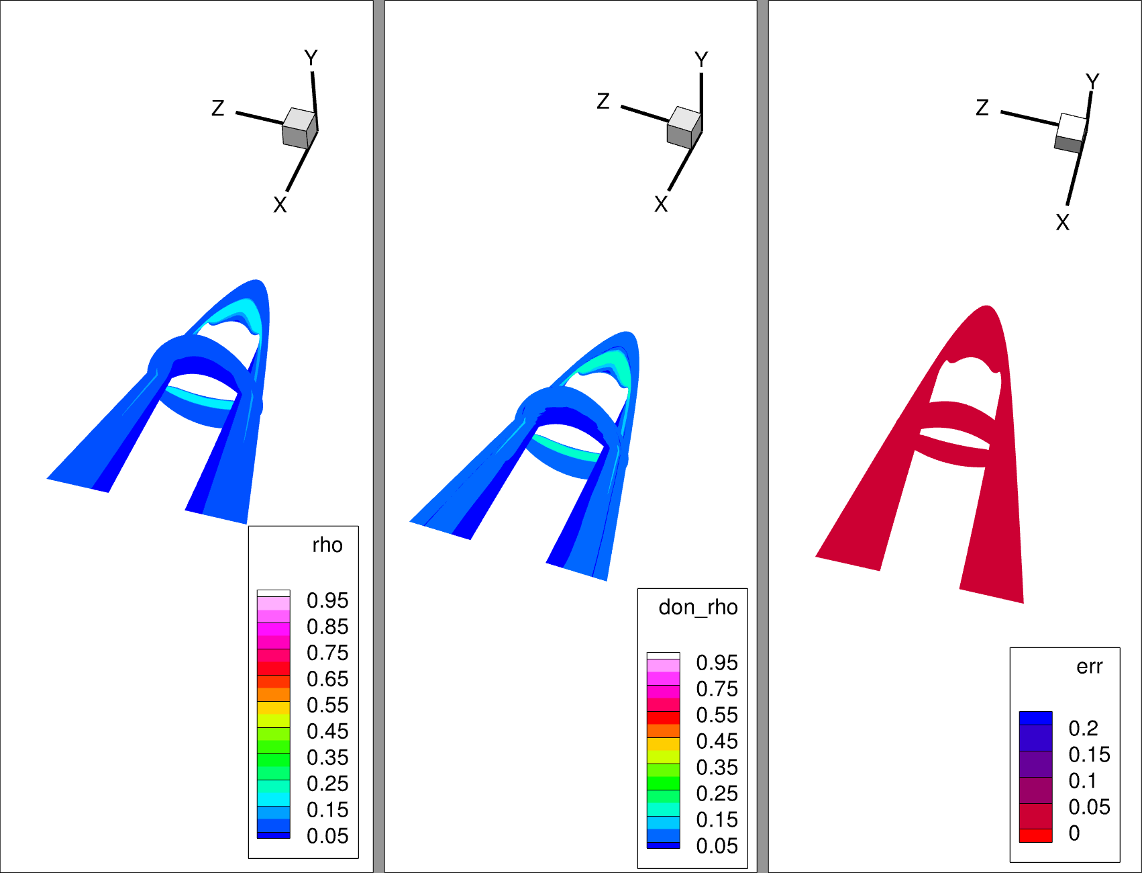}
  \caption{Density ($\rho$) distributions of the fluid domain around an AEDC waverider for an unseen 3$^\circ$ angle of attack. From left to right the first three columns are: the flow field predicted by the US3D solver; the flow field  predicted by the surrogate two-step DeepONet; and the pointwise error (difference) between the US3D solve and DeepONet flow field predictions across the entire volume. The last three columns are slices of the density distribution from left to right: the flow field predicted by the US3D solver; the flow field  predicted by the surrogate two-step DeepONet; and the pointwise absolute error (absolute difference) between the US3D solve and DeepONet flow field predictions. }
  \label{fig:vol}
\end{figure}

In Figure \ref{fig:vol}, the density ($\rho$) distribution at a 3$^\circ$ angle of attack is shown for the entire fluid domain encompassing the AEDC waverider. The first, second and third column in  in Figure \ref{fig:vol} illustrate the density simulated by the US3D solver, predicted by the two-step DeepONet, and the pointwise error, respectively. However, in the fourth and fifth columns, we present density slices in the XY and YZ planes obtained from the US3D solver and DeepONet, respectively. In the sixth column of \autoref{fig:vol}, we show the pointwise absolute error between the density slices obtained from US3D and DeepONet, respectively. Notably, as depicted in \autoref{fig:vol}, the two-step DeepONet effectively captures the shock location in the flow fields. The global relative error between the numerical and predicted density is 5.1\%. Additionally, the inference time for predicting the flow field using the trained DeepONet is 1.032 ms on an A100 NVIDIA GPU, significantly faster than the wall time taken by the US3D solver, which amounts to 32,000 core hours.

In \autoref{hyper_table}, we show all the hyper-parameters used in generating the results shown in \autoref{fig:harv_rho}, \autoref{fig:3D_comparison}  and \autoref{fig:vol}.  

\begin{table}[h]
\centering
\caption{Hyperparameters for the experiments in this study, referenced by figure number.}

\begin{tabular}{ |p{2cm}||p{4cm}|p{5cm}|p{2cm}|  }
 \hline
 \textbf{Figure No.} & \textbf{No. of Hidden Layers} & \textbf{No. of Neurons in each layer} & \textbf{Activation function}\\
 \hline
 \autoref{fig:harv_rho}   & Branch Net = 2, Trunk Net=2  & Branch Net = 100, Trunk Net=100  & Tanh\\
 \hline
 \autoref{fig:3D_comparison}&   Branch Net = 2, Trunk Net=3 & Branch Net = 64, Trunk Net=64  & ReLu \\
 \hline
 \autoref{fig:vol} & Branch Net = 8, Trunk Net=8 & Branch Net = 48, Trunk Net=48  & ReLu \\\hline
\end{tabular}
\label{hyper_table}
\end{table}


\section{Summary and Conclusions}
We have successfully developed surface and volume DeepONet based surrogate models of the AEDC Waverider under hypersonic flow conditions. To address capturing the position of shocks and other areas of high gradients, we use the two-step training method for DeepONets. The two-step method ensures the selection of appropriate basis functions, resulting in improved convergence and accuracy. It offers a reduced training cost advantage compared to the vanilla DeepONet, as it separately trains the branch and trunk networks. We empirically validate the effectiveness of DeepONets in maintaining sufficient accuracy in flow field representation while significantly enhancing computational efficiency compared to using traditional CFD solvers. This enhancement makes DeepONets suitable for various engineering applications, including shape optimization. 

Crucially, DeepONets demonstrate minimal generalization error across the dataset, enabling accurate prediction of the flow field with approximately a 32,000-fold speed-up compared to the CFD baseline. However, mitigating the computational complexity associated with training a DeepONet can be achieved by seamlessly extending the training procedures across multiple GPUs in a data-parallel sense, as outlined in \cite{goyal2017accurate}. The framework presented here is versatile and capable of tackling more intricate issues involving multiple inputs, such as varying Mach numbers and diverse parameterizations of geometry, which can be fed into either the branch or trunk networks. Consequently, with minor adjustments, this framework can accommodate optimization in high-speed flow regimes characterized by flow unsteadiness, shocks, non-equilibrium chemistry, and even adaptive geometry. The future work includes integration of CFD surrogate with optimization packages such as Dakota for concept evaluation. 


\section*{Acknowledgments}
Copyright 2024 Carnegie Mellon University, Khemraj Shukla, Luis Bravo, Nicholas Plewacki, Anindya Ghoshal, George Karniadakis. This material is based upon work funded and supported by the Department of Defense under Contract No. FA8702-15-D-0002 with Carnegie Mellon University for the operation of the Software Engineering Institute, a federally funded research and development center, and by DEVCOM Army Research Laboratory grants, 185 W911NF-19-1-0225 and W911NF-22-2-0058.
 
[DISTRIBUTION STATEMENT A] This material has been approved for public release and unlimited distribution. 
 
The authors gratefully acknowledge the High-Performance Computing Modernization Program (HPCMP) resources and support provided by the Department of Defense Supercomputing Resource Center (DSRC) as part of the 2022 Frontier Project, Large-Scale Integrated Simulations of Transient Aerothermodynamics in Gas Turbine Engines. 
 
The views and conclusions contained in this document are those of the authors and should not be interpreted as representing the official policies or positions, either expressed or implied, of the DEVCOM Army Research Laboratory or the U.S. Government. 
 
This work is licensed under a Creative Commons Attribution-Non Commercial 4.0 International License. 
 
Requests for permission for non-licensed uses should be directed to the Software Engineering Institute at permission@sei.cmu.edu.
 
The U.S. Government is authorized to reproduce and distribute reprints for Government purposes notwithstanding any copyright notation herein.
DM23-2250
\bibliography{refs/00main.bib}

\end{document}